------------------------ RevTex file ------------------------
\documentstyle[version2,aps]{revtex}
\begin{document}
\draft
\twocolumn
\widetext
\begin{title}
Excitation Spectra of the Negative-$U$ Hubbard Model:\\
A Small-Cluster Study
\end{title}
\author{Y. Ohta,\cite{byline1} A. Nakauchi, R. Eder,\cite{byline2}
K. Tsutsui, and S. Maekawa}
\begin{instit}
Department of Applied Physics, Nagoya University,
Nagoya 464-01, Japan
\end{instit}
\receipt{13 March 1995}
\begin{abstract}
An exact-diagonalization technique on small clusters is used to
study low-lying excitations and superconductivity in the
two-dimensional negative-$U$ Hubbard model.
We first calculate the Bogoliubov-quasiparticle spectrum,
condensation amplitude, and coherence length as functions of
the coupling strength $U/t$, thereby working out how the picture
of Bogoliubov quasiparticles in the BCS superconductors is affected
by increasing the attraction.
We then define the Cooper-pair operator as a spatially-extended
composite boson and make a variational evaluation of its internal
structure.  We thereby calculate the single-particle spectral
function of the Cooper pair and obtain the dispersion relation
for its translational motion.  The dynamical density
correlation function of the pairs is also calculated.
We thus demonstrate the applicability of our numerical method
for gaining insight into low-lying excitations of models for the
intermediate coupling superconductivity relevant to cuprate
materials.
\end{abstract}
\pacs{74.20.Mn, 75.40.Mg, 71.27.+a}
\narrowtext
\topskip11cm

\section{INTRODUCTION}

High-temperature superconductors typically show very short
coherence length and small carrier numbers, and may be in
an intermediate regime between two well-understood regimes,
i.e., the BCS weak-coupling superconductivity and
Bose-condensation of pre-formed bosons.  Although much theoretical
effort has been devoted to examining this crossover regime,
not very much is known, in particular, of the low-lying
excitations [1,2].  To consider this problem, we have studied
the negative-$U$ Hubbard model because of its controllable
strength of attraction.  This model, much discussed recently
[3--10], may be a good reference system to more relevant models,
such as the $t$$-$$J$ model, for cuprate superconductivity.

The negative-$U$ Hubbard Hamiltonian is written
\begin{equation}
H=-t\sum_{<{\bf ij}>\sigma}
   (c^\dagger_{{\bf i}\sigma}c_{{\bf j}\sigma}+{\rm H.c.})
  + U \sum_{\bf i}n_{{\bf i}\uparrow}n_{{\bf i}\downarrow}
\label{Eq1}
\end{equation}
with a negative value of $U$, where $c^\dagger_{{\bf i}\sigma}$
($c_{{\bf i}\sigma}$) is the electron creation (annihilation)
operator at site ${\bf i}$ and spin $\sigma$
(=$\uparrow,\downarrow$), and
$n_{{\bf i}\sigma}$$=$$c^\dagger_{{\bf i}\sigma}
c_{{\bf i}\sigma}$.
The summation $<$${\bf ij}$$>$ is taken over all the
rearest-neighbor pairs on the two-dimensional square lattice.
We employ an exact-diagonalization technique on small clusters;
a $4$$\times$$4$ cluster with periodic boundary condition is
used throughout the work, and ground states and dynamical
correlation functions are calculated by the Lanczos algorithm.

As an approach from weak-attraction limit we first calculate
the Bogoliubov quasiparticle spectra in the BCS pairing theory
as a function of $U/t$, and demonstrate how the picture of
Bogoliubov quasiparticles is affected by increase of the strength
of attraction [11].  Condensation amplitude and coherence
length are also estimated.  We then make a variational calculation
of the Cooper-pair wave function and examine $U/t$ dependence of
the internal structure of the pair.
Next we calculate the wave functions of pairs with non-zero
momenta and examine the single-particle excitation spectra of the
pair to see its translational motion.
Finally we calculate the dynamical pair-density correlation,
and examine a density fluctuation of the pairs.
We thus demonstrate the applicability of our numerical method
for gaining insight into low-lying excitations of models for the
intermediate coupling superconductivity.

\topskip0cm

\section{BOGOLIUBOV QUASIPARTICLE}

Let us first examine the validity of the Bogoliubov-quasiparticle
picture in the negative-$U$ Hubbard model as a function of
attraction $U/t$.  We use a recently proposed technique [12],
i.e., an exact calculation of Bogoliubov quasiparticle spectrum
on small clusters.  We thereby see whether low-lying states of the
negative-$U$ Hubbard cluster of a given attraction $U/t$ can be
described by this picture.

We define the one-particle anomalous Green's function as
\begin{equation}
G({\bf k},z)=
\langle\psi^{N+2}_0|c^\dagger_{{\bf k}\uparrow}
{1 \over {z-H+E_0}}
c^\dagger_{-{\bf k}\downarrow}|\psi^{N}_0\rangle
\label{Eq2}
\end{equation}
where $|\psi^N_0\rangle$ is the cluster ground state with
$N$ (even) electrons, $E_0$ is the ground-state energy
(averaged over energies of the $N$ and $N$$+$$2$ electron
states), and $c^\dagger_{{\bf k}\sigma}$ is the Fourier
transform of $c^\dagger_{{\bf i}\sigma}$.  We then define the
spectral function $F({\bf k},\omega)$$=$$-(1/\pi){\rm Im}
G({\bf k},\omega$$+$$i\eta)$ with $\eta$$=$$+0$ and
its frequency integral
\begin{equation}
F_{\bf k}=\langle
\psi^{N+2}_0|c^\dagger_{{\bf k}\uparrow}
c^\dagger_{-{\bf k}\downarrow}|\psi^N_0\rangle.
\label{Eq3}
\end{equation}
This Green's function should describe the excitation
of a Bogoliubov quasiparticle in the cluster, i.e.,
$F({\bf k},\omega)$$=$$F_{\bf k}\delta (\omega$$-$$E_{\bf k})$
with $F_{\bf k}$$=$$\Delta_{\bf k}/2E_{\bf k}$ for the
quasiparticle energy $E_{\bf k}$ and gap function
$\Delta_{\bf k}$, provided that the low-lying states of the
cluster can be described by the microcanonical version of the
BCS pairing theory [13].

The calculated results for $F({\bf k},\omega)$ at $N$$=$$8$ and
$10$ are shown in Figs.~1 (a) and (b), respectively.
Fermi momentum ${\bf k}_{\rm F}$ is at $(\pi/2,0)$ for $N$$=$$8$
and at $(\pi/2,\pi/2)$ and $(\pi,0)$ for $N$$=$$10$.  We find that,
at $U/t$$=$$-2$, the spectrum fits fairly well with the BCS
spectrum (dotted curves) with $s$-wave gap function
$\Delta_{\bf k}$$=$$\Delta_0$:  pronounced low-energy peaks
appear at ${\bf k}_{\rm F}$, smaller peaks appear at higher
energies for other momenta.  The observed gap is isotropic, and
the spectral weights are consistent with the BCS form of
the condensation amplitude $F_{\bf k}$ with a maximum at
${\bf k}_{\rm F}$ (see Fig.~2).  However, at $U/t$$=$$-6$, the
peaks already extend to higher-energy regions and the dispersion
becomes rather deformed, indicating that the Bogoliubov
quasiparticle is becoming a less well-defined excitation although
the lowest-energy peaks at ${\bf k}_{\rm F}$ are still sharp and
well-defined.  For much larger values of $|U|/t$ the spectra
are totally incoherent and the notion of Bogoliubov quasiparticles
loses its significance.

We also calculate the coherence length $\xi$ defined as
\begin{equation}
\xi^2=\sum_{\bf k}|\nabla_{\bf k}F_{\bf k}|^2
/\sum_{\bf k}|F_{\bf k}|^2.
\label{Eq4}
\end{equation}
The result (see Fig.~2) shows a rapid but smooth crossover
from a Cooper-paired state to a Bose condensed state of tightly
bound pairs.  At $|U|/t$$\simeq$$3$$-$$4$, $\xi$ is of the
size of the nearest-neighbor to next-nearest-neighbor distance.
Note that $F_{\bf k}$ goes to zero when $|U|/t$$\rightarrow$$0$
for ${\bf k}$$\ne$${\bf k}_{\rm F}$ in agreement with the BCS
pairing theory; nondivergence of $\xi$ at $|U|/t$$\rightarrow$$0$
is an obvious finite-size effect.

\section{COOPER PAIR}

As an approach from strong-attraction regime we introduce a
spatially-extended `composite-boson' operator defined as
$b_{\bf i}^\dagger = \sum_{\bf r}\alpha({\bf r})
c_{{\bf i}\uparrow}^\dagger
c_{{\bf i}+{\bf r}\downarrow}^\dagger$
[14] and its Fourier transform $b_{\bf q}^\dagger$ with
$\alpha({\bf k})=\sum_{\bf r}\alpha({\bf r})
e^{i{\bf k}\cdot{\bf r}}$.
We then define a variational wave function by adding a
zero-momentum pair of up and down electrons to the $N$-electron
ground state:
\begin{eqnarray}
|{\tilde \psi}^{N+2}_0\rangle
&=&{\rm const.}\times b^\dagger_{{\bf q}=0}|\psi^N_0\rangle
\nonumber \\
&=&{\rm const.}\times{1\over\sqrt{N_s}}\sum_{\bf k}\alpha({\bf k})
c_{-{\bf k}\uparrow}^\dagger
c_{{\bf k}\downarrow}^\dagger
|\psi^N_0\rangle
\label{Eq5}
\end{eqnarray}
where $N_s$ is the number of lattice sites.
We determine variational coefficients $\alpha({\bf k})$
so as to maximize the overlap with the exact wavefunction
$|\langle\psi^{N+2}_0|{\tilde \psi}^{N+2}_0\rangle|^2$
under the condition that $|{\tilde \psi}^{N+2}_0\rangle$
be normalized to unity,
which leads to a generalized eigenvalue problem
\begin{equation}
\sum_{{\bf k}'}[F_{\bf k}^*F_{{\bf k}'}-
\lambda N_{{\bf kk}'}]\alpha({\bf k}')=0
\label{Eq6}
\end{equation}
with $F_{\bf k}$ of Eq.~3 and
$N_{{\bf kk}'}$$=$$\langle\psi^N_0|
c_{{\bf k}\downarrow}c_{-{\bf k}\uparrow}
c^\dagger_{-{\bf k}'\uparrow}c^\dagger_{{\bf k}'\downarrow}
|\psi^N_0\rangle$ [15].
We find for the cluster that an overlap of $\lambda$$\simeq$$0.95$
is generally achieved.
Defining ${\tilde F}_{\bf k}$ by replacing $\psi^{N+2}_0$ with
${\tilde \psi}^{N+2}_0$ in Eq.~3, we have
${\tilde F}_{\bf k}$$=$$\sum_{{\bf k}'}
\alpha^*({\bf k}')N_{{\bf kk}'}$
and $\sum_{\bf k}{\tilde F}_{\bf k}\alpha({\bf k})$$=$$1$.
One may use these relations with the exact $F_{\bf k}$ to check
the validity of the variational wave function; actually we find
values of $\sum_{\bf k}F_{\bf k}\alpha({\bf k})$
($=$$\lambda^{1/2}$) to be close to 1 within a few percent.

Figure 3 shows the Cooper-pair wave function thus determined;
the pair is added to the state with $N$$=$$8$ and
${\bf k}_{\rm F}$$=$$(\pi/2,0)$.  We find in Fig.~3 (a) that
when $|U|/t$$\alt$$6$ $\alpha({\bf k})$ has a peak at
${\bf k}_{\rm F}$, and with increasing $|U|/t$, it broadens
over the entire Brillouin zone; the Pauli principle acting
between the added electrons and Fermi sea plays an essential
role in the structure of the Cooper-pair wave function.
In real space, as is expected from the BCS pairing theory,
$\alpha({\bf r})$ (see Fig.~3 (b)) shows an oscillation of
wavelength $2\pi/k_{\rm F}$ corresponding to the peak in
$\alpha({\bf k})$, and the oscillation decays with the length
scale of $\xi$ as one may see via $U/t$ dependence of
$\alpha({\bf r})$.

\section{EXCITATIONS OF COMPOSITE BOSONS}

Let us next examine excitation spectra for translational motion
of the Copper pair.  We first introduce a creation operator of
the finite-momentum (${\bf q}$) pair which is defined as
\begin{equation}
b_{\bf q}^\dagger = {1\over\sqrt{N_s}}\sum_{\bf k}
\alpha_{\bf q}({\bf k})c_{-{\bf k}+{\bf q}\uparrow}^\dagger
c_{{\bf k}\downarrow}^\dagger
\label{Eq7}
\end{equation}
with ${\bf q}$-dependent internal structure
$\alpha_{\bf q}({\bf k})$ (which reduces to $b^\dagger_{\bf q}$
defined in Sec.~III at ${\bf q}$$=$$0$), and again perform
the variational evaluation of $\alpha_{\bf q}({\bf k})$ by a
generalization of the above procedure where $F_{\bf k}$ and
$N_{{\bf kk}'}$ also depend on ${\bf q}$.  The operator thus
obtained is compared below with the on-site pair-field operator
$b_{\bf q}^\dagger$$=$$(1/\sqrt{N_s})\sum_{\bf i}
c_{{\bf i}\uparrow}^\dagger
c_{{\bf i}_\downarrow}^\dagger
e^{-i{\bf q}\cdot{\bf i}}$,
i.e., Eq.~(7) with $\alpha_{\bf q}({\bf k})$$=$$1$.
We find for the cluster with $N$$=$$8$ that the overlap
$\lambda$$\agt$$0.7$ (mostly around $\sim$$0.8$) is achieved
by the optimization; the presence of the Fermi sea is again
found to play an important role in the ${\bf k}$ and ${\bf q}$
dependence of $\alpha_{\bf q}({\bf k})$.

We then define the pair-addition spectrum as
\begin{equation}
P({\bf q},\omega)=-{1\over\pi}\Im\langle\psi^N_0|b_{\bf q}
{1\over{\omega+i\eta-H+E^N_0}}b^\dagger_{\bf q}|\psi^N_0\rangle
\label{Eq8}
\end{equation}
whereby examining the single-particle excitation of the Cooper
pair.  The chemical potential for the Cooper pair is defined as
$\mu_p$$=$$\partial E/\partial N_p$$=$$E^{N+2}_0-E^N_0$ where
$N_p$ ($=$$N/2$) is the number of pairs.
Figure 4 shows the calculated result for $P({\bf q},\omega)$ with
optimized values of $\alpha_{\bf q}({\bf k})$; we compare this
with the result obtained for the on-site pair-field operator.
We first of all find that with the optimization higher-energy
peaks are suppressed strongly and the lowest-energy peak is
enhanced at each momentum ${\bf q}$, indicating that the
spatially-extended `composite boson' describes the low-energy
excitation of the system fairly well [16]; its effective mass,
e.g., is noted to become heavier with increasing $|U|/t$.
An indication of the off-diagonal long-range order is also seen
in the calculated momentum distribution
$\langle b^\dagger_{\bf q}b_{\bf q}\rangle$, i.e., its enhancement
at ${\bf q}$$=$$(0,0)$, although in the thermodynamic
limit a macroscopic number of bosons are condensed into
${\bf q}$$=$$0$ state.
One may then check how the Bogoliubov theory for interacting
bosons [17] works for these composite operators; we directly
calculate the density correlation function defined as
\begin{equation}
B({\bf q},\omega)=-{1\over \pi}\Im\langle\psi^N_0|d_{-{\bf q}}
{1\over \omega+i\eta-H+E_0^N}d_{\bf q}|\psi^N_0\rangle
\label{Eq9}
\end{equation}
with the density operator $d_{\bf q}$$=$$(1/\sqrt{N_s})
\sum_{\bf k}b^\dagger_{{\bf k}+{\bf q}}b_{\bf k}$.
The calculated result for $B({\bf q},\omega)$ at $U/t$$=$$-4$
is shown in Fig.~5; the spectrum is approximated well by a single
peak around ${\bf q}$$=$$(0,0)$, which is consistent with the
expected picture of the collective sound mode.  For larger
momenta this mode decays into pair-breaking excitations
observed in $F({\bf k},\omega)$.

\section{SUMMARY}

We have proposed numerical techniques for examining the low-lying
excitation spectra of the two-dimensional negative-$U$ Hubbard
model and studied the
(i) Bogoliubov-quasiparticle excitation,
(ii) condensation amplitude and coherence length,
(iii) internal structure of the Cooper pairs, and
(iv) single-particle and density excitations of the pairs,
as functions of the attractive interaction.
We have demonstrated how the picture of Bogoliubov-quasiparticle
excitations in the BCS superconductors loses its significance
as the coupling strength increases and how the picture of
spatially-extended composite bosons works as the coupling strength
decreases.  Application of the present numerical technique to
strongly-correlated electron models will help us gain insight
into the intermediate-coupling superconductivity relevant to
cuprate materials.

\acknowledgments
This work was supported by Priority-Areas Grants from
the Ministry of Education, Science, and Culture of
Japan.  R. E. acknowledges financial support by the
Japan Society for Promotion of Science.  Computations
were partly carried out in the Computer Center of
Institute for Molecular Science, Okazaki National
Research Institutes.


\figure{Bogoliubov quasiparticle spectra $F({\bf k},\omega)$
for the $4$$\times$$4$ cluster with filling of (a) $N$$=$$8$
at $U/t$$=$$-2$ (left panel) and $-6$ (right panel), and (b)
$N$$=$$10$ at $U/t$$=$$-2$ (left panel) and $-6$ (right panel).
Dotted curves in the left panels show the BCS spectral function
obtained for (a) $\Delta_0/t$$=$$0.28$ and (b) $0.4$.  We use
the value $\eta/t$$=$$0.15$ for Lorentzian broadening of the
spectra.
\label{fig1}}
\figure{Condensation amplitude $F_{\bf k}$ and coherence length
$\xi$ as a function of the attraction $U/t$ calculated for the
$4\times 4$ cluster with filling of $N$$=$$8$.
\label{fig2}}
\figure{(a) Cooper-pair wave function $\alpha({\bf k})$ and
(b) its Fourier transform $\alpha({\bf r})$ calculated for the
$4$$\times$$4$ cluster with filling of $N$$=$$8$.
\label{fig3}}
\figure{Pair-addition spectra $P({\bf q},\omega)$ calculated
for the $4$$\times$$4$ cluster with filling of $N$$=$$8$ at
$U/t$$=$$-4$.  Vertical dotted lines indicate $\mu_p$.
Left panel: the spectra for $b_{\bf q}^\dagger$
with optimized $\alpha_{\bf q}({\bf k})$ values.
Right panel: the spectra for the on-site pair-field operator.
Lorentzian broadening of $\eta/t$$=$$0.05$ is used.
\label{fig4}}
\figure{Density correlation function $B({\bf q},\omega)$ of
composite bosons calculated for the $4$$\times$$4$ cluster with
$N$$=$$4$ and $U/t$$=$$-4$.
Lorentzian broadening of $\eta/t$$=$$0.05$ is used.
\label{fig6}}

\begin{references}
\bibitem[*]{byline1} Present address: Department of Physics,
                     Faculty of Science, Chiba University,
                     Inage-ku, Chiba 263, Japan.
\bibitem[**]{byline2} Present address: Department of Applied and
                      Solid State Physics, University of Groningen,
                      Nijenborgh 4, 9747 AG Groningen,
                      The Netherlands.
\bibitem{[1]} A. J. Leggett, in {\it Modern Trends in the Theory
              of Condensed Matter} edited by A. Pekalski and
              J. Przystawa (Springer-Verlag, Berlin, 1980), p.~13;
              J. Phys. (Paris) {\bf 41}, C7-19 (1980).
\bibitem{[2]} P. Nozier\`es and S. Schmitt-Rink,
              J. Low Temp. Phys. {\bf 59}, 195 (1985).
\bibitem{[3]} See, e.g., a review article by
           R. Micnas, J. Ranninger, and S. Robaszkiewicz,
           Rev. Mod. Phys. {\bf 62}, 113 (1990).
\bibitem{[4]} R. T. Scalettar, E. Y. Loh, J. E. Gubernatis,
           A. Moreo, S. R. White, D. J. Scalapino, R. L. Sugar,
           and E. Dagotto,
           Phys. Rev. Lett. {\bf 62}, 1407 (1989).
\bibitem{[5]} A. Moreo and D. J. Scalapino, Phys. Rev. Lett.
           {\bf 66}, 946 (1991).
\bibitem{[6]} A. Moreo, D. J. Scalapino, and S. R. White, Phys. Rev.
           B {\bf 45}, 7544 (1992).
\bibitem{[7]} M. Randeria, N. Trivedi, A. Moreo, and R. T. Scalettar,
           Phys. Rev. Lett. {\bf 69}, 2001 (1992).
\bibitem{[8]} J. O. Sofo, C. A. Balseiro, and H. E. Castillo,
           Phys. Rev. B {\bf 45}, 9860 (1992).
\bibitem{[9]} H. E. Castillo and C. A. Balseiro, Phys. Rev. B
           {\bf 45}, 10549 (1992).
\bibitem{[10]} F. F. Assaad, W. Hanke, and D. J. Scalapino,
           Phys. Rev. Lett. {\bf 71}, 1915 (1993) and
           Phys. Rev. B {\bf 49}, 4327 (1994).
\bibitem{[11]} Y. Ohta, A. Nakauchi, R. Eder, K. Tsutsui, and
           S. Maekawa, Physica C {\bf 235-240}, 2169 (1994).
\bibitem{[12]} Y. Ohta, T. Shimozato, R. Eder, and S. Maekawa,
           Phys. Rev. Lett. {\bf 73}, 324 (1994).
\bibitem{[13]} J. R. Schrieffer,
              {\it Theory of Superconductivity}
              (Benjamin Inc., New York, 1964).
\bibitem{[14]} Bosonic commutation relations are not exactly
satisfied reflecting Fermionic nature of the operator although
we use the term `composite boson' by convention: see Ref.~[3].
Also, the singlet pairing operator
$$
b_{\bf i}=\alpha(0)c_{{\bf i}\uparrow}
c_{{\bf i}\downarrow}+
{1\over\sqrt{2}}\sum_{{\bf r}\ne 0}\alpha({\bf r})
(c_{{\bf i}\uparrow}c_{{\bf i}+{\bf r}\downarrow}-
c_{{\bf i}\downarrow}c_{{\bf i}+{\bf r}\uparrow})
$$
may be used but the results are not different because
only the singlet $s$-wave pairing is relevant in the
present model.
\bibitem{[15]} Mininization of the total energy leads to
              a generalized eigenvalue equation
$\sum_{{\bf k}'} [H_{{\bf kk}'}-EN_{{kk}'}]\alpha({\bf k}')$$=$$0$
with $H_{{\bf kk}'}$$=$$\langle\psi^N_0|
c_{{\bf k}\downarrow}c_{-{\bf k}\uparrow} H
c^\dagger_{-{\bf k}'\uparrow}c^\dagger_{{\bf k}'\downarrow}
|\psi^N_0\rangle$,
which may also be used to find values of $\alpha({\bf k})$,
although we find the results are very similar.
\bibitem{[16]} A single peak is found at ${\bf q}$$=$$(\pi,\pi)$.
This is because the state $b_{\bf q}^\dagger |\psi_0^N\rangle$
is an eigenstate of the Hamiltonian, i.e.,
$[H,b^\dagger_{\bf q}]=-Ub^\dagger_{\bf q}$
with ${\bf q}=(\pi,\pi)$.
See S. Zhang, Phys. Rev. Lett. {\bf 65}, 120 (1990).
\bibitem{[17]} N. N. Bogoliubov, J. Phys. USSR {\bf 11},
               23 (1947).
\end{references}
\end{document}